\documentstyle[12pt]{article}
\let\ssection=\section
\renewcommand{\section}{\setcounter{equation}{0}\ssection}

\newcommand{\be}{\begin{enumerate}}
\newcommand{\ee}{\end{enumerate}}
\newcommand{\bi}{\begin{itemize}}
\newcommand{\ei}{\end{itemize}}
\newcommand{\beq}{\begin{equation}}
\newcommand{\eeq}{\end{equation}}
\newcommand{\beqa}{\begin{eqnarray}}
\newcommand{\eeqa}{\end{eqnarray}}

\newcommand{\eq}[1]{(\ref{#1})}
\newcommand{\eqs}[2]{(\ref{#1}--\ref{#2})}
\newcommand{\ie}{{\em i.e.,\ }}

\renewcommand{\a}{\alpha}                   
\renewcommand{\b}{\beta}                    
\newcommand{\g}{\gamma}

\newcommand{\s}{\sigma}

\newcommand{\bra}[1]{\langle #1 |}
\newcommand{\braket}[2]{\bra{#1} #2\rangle}
\newcommand{\ket}[1]{| #1 \rangle}

\newcommand{\tr}{{\rm\, tr}}

\newcommand\mathC{\mkern1mu\raise2.2pt\hbox{$\scriptscriptstyle|$}
                {\mkern-7mu\rm C}}
\newcommand{\mathR}{{\rm I\! R}}                

\newcommand{\C}{{\tilde{C}}}
\newcommand{\D}{{\cal D}}
\renewcommand{\P}{{\cal P}}
\newcommand{\PV}{{\cal P}({\cal V})}
\newcommand{\UP}{{\cal UP}}
\renewcommand{\H}{{\cal H}}
\newcommand{\V}{{\cal V}}

\newcommand{\h}[1]{({#1}_{t_1},{#1}_{t_2},\ldots,{#1}_{t_n})}

\begin{document}

\begin{titlepage}
\hspace{8truecm} Imperial/TP/93--94/42

\hspace{8truecm} DAMTP 94--44

\begin{center}
   {\large\bf The Classification of Decoherence Functionals:\\[0.2truecm]
		 An Analogue of Gleason's Theorem}
\end{center}
\smallskip
\begin{center}
        C.J.~Isham\footnote{Permanent address: Blackett
        Laboratory, Imperial College, South Kensington, London SW7
        2BZ; email: c.isham@ic.ac.uk}\\[0.2cm]
        Isaac Newton Institute for Mathematical Sciences\\
        University of Cambridge\\
        20 Clarkson Road\\
        Cambridge CB3 0EH\\
\begin{center} and \end{center}
        N.~Linden\footnote{email: n.linden@newton.cam.ac.uk}\\[0.2cm]
        D.A.M.T.P.\\
        University of Cambridge\\
        Cambridge CB3 9EW\\
\begin{center} and \end{center}
		S.~Schreckenberg\\[0.2cm]
		Blackett Laboratory\\
		Imperial College\\
		South Kensington\\
		London SW7 2BZ
\end{center}

\begin{center} June 1994\end{center}
\vfill\eject
\begin{abstract}
 Gell-Mann and Hartle have proposed a significant generalisation of
quantum theory with a scheme whose basic ingredients are `histories'
and decoherence functionals. Within this scheme it is natural to
identify the space $\UP$ of propositions about histories with an
orthoalgebra or lattice. This raises the important problem of
classifying the decoherence functionals in the case where $\UP$ is the
lattice of projectors $\PV$ in some Hilbert space $\V$; in effect we
seek the history analogue of Gleason's famous theorem in standard
quantum theory. In the present paper we present the solution to this
problem for the case where $\V$ is finite-dimensional. In particular,
we show that every decoherence functional $d(\a,\b)$, $\a,\b\in\PV$
can be written in the form $d(\a,\b)=\tr_{\V\otimes\V}(\a\otimes\b X)$
for some operator $X$ on the tensor product space $\V\otimes\V$.
 \end{abstract}

\end{titlepage}

\section{Introduction}
\label{Sec:intro}
 There has been much interest in the `consistent-histories'
approach to quantum theory following the fundamental work of Griffiths
\cite{Gri84}, Omn\`es \cite{Omn88a,Omn88b,Omn88c,Omn89,Omn90,Omn92},
and Gell-Mann and Hartle \cite{GH90a,GH90b,GH90c,Har91a,Har91b,GH92}.
This approach is motivated in part by a result in conventional quantum
theory concerning measurements of a time-ordered sequence of
properties $\a:=\h{\a}$ with $t_1<t_2<\cdots<t_n$
(what we shall call a {\em homogeneous history\/} of the system).
Namely, if the state at some initial time $t_0$ is the density matrix
$\rho_{t_0}$, then the joint probability of finding all the properties is
 \beq
        {\rm Prob}(\a_{t_1},\a_{t_2},\ldots,\a_{t_n};\rho_{t_0})=
            \tr_\H(\C_\a^\dagger\rho_{t_0} \C_\a)      \label{Prob:a1-an}
\eeq
where the `class' operator $\C_\a$ is defined in terms  of the
Schr\"odinger-picture projection operators $\a_{t_i}$ as
 \beq
    \C_\a:=U(t_0,t_1)\a_{t_1} U(t_1,t_2)\a_{t_2}\ldots U(t_{n-1},t_n)
        \a_{t_n}U(t_n,t_0)=
        \a_{t_1}(t_1)\a_{t_2}(t_2)\ldots\a_{t_n}(t_n) \label{Def:C_a}
 \eeq
where $\a_{t_i}(t_i):= U(t_0,t_i)\a_{t_i} U(t_0,t_i)^\dagger$ is
the Heisenberg picture operator defined with respect to the fiducial
time $t_0$, and $U(t,t')=e^{-i(t-t')H/\hbar}$ is the usual
time-evolution operator on the Hilbert space $\H$ of the system.

    The main assumption of the consistent-histories interpretation of
quantum theory is that, under appropriate `consistency' conditions,
the probability assignment \eq{Prob:a1-an} is still meaningful for a
{\em closed\/} system where there are no external observers to produce
a measurement-induced reduction of the state-vector. The satisfaction
of these conditions is determined by the values of the decoherence
functional $d_{(H,\rho)}(\a,\b)$ defined by
 \beq
    d_{(H,\rho)}(\a,\b)=\tr(\C_\a^\dagger\rho_{t_0}\C_\b)    \label{Def:d}
 \eeq
 where $\a=\h{\a}$ and $\b=(\b_{t_1'},\b_{t_2'},\ldots,\b_{t_m'})$ are
an arbitrary pair of homogeneous histories. Note that, as suggested by
the notation $d_{(H,\rho)}$, both the initial state and the dynamical
structure (\ie the Hamiltonian $H$) are coded in the decoherence
functional: in our approach, a history $\h{\a}$ itself is just a
`passive', time-ordered sequence of propositions.

    An important feature of the work of Gell-Mann and Hartle is their
supposition that this formalism can be extended to include disjoint
sums of homogeneous histories, known as {\em inhomogeneous\/}
histories. In particular, if $\a,\b$ are `disjoint',
the class operator for the sum $\a\oplus\b$
(interpreted heuristically as ``$\a$ {\em or\/} $\b$'') is assumed to
be $\C_{\a\oplus\b}=\C_\a+\C_\b$. Likewise, there is a negation
operation $\neg$ with $\C_{\neg\a}=1-\C_\a$.

	Gell-Mann and Hartle have suggested a major development of
this scheme to one in which a history is regarded as a fundamental
entity in its own right, not necessarily just a  time-ordered sequence
of projection operators \cite{Har93a}. The physical results are
obtained from the values of a decoherence functional, defined now
as a complex-valued function of pairs of histories that satisfies certain
conditions suggested naturally by the form \eq{Def:d} of standard
quantum theory. It is hoped that a framework of this type will be of
value in tackling the quantum theory of gravity, especially the
infamous `problem of time'.

    In \cite{Ish94a} and \cite{IL94a} it was suggested that the
natural mathematical tools with which to discuss theories of this type
are the algebraic structures employed in quantum logic. Specifically,
the set of all histories (or, more precisely, the set of all {\em
propositions\/} about histories or possible universes) can be modelled
by an orthoalgebra $\UP$ whose triad of operations
$(\oplus,\neg,\leq)$ correspond respectively to the disjoint sum,
negation, and coarse-graining operations postulated by  Gell-Mann and
Hartle. This suggestion is motivated by the demonstration in
\cite{Ish94a} that the set of all history-propositions in standard
quantum theory can indeed be identified with the lattice of a certain
Hilbert space. The key idea is to associate the homogeneous history
proposition $\h{\a}$ with the operator
$\a_{t_1}\otimes\a_{t_2}\otimes\cdots\otimes\a_{t_n}$ which is a
genuine {\em projection\/} operator on the tensor product
$\H_{t_1}\otimes\H_{t_2}\otimes\cdots\otimes\H_{t_n}$ of $n$ copies of
the Hilbert space $\H$ on which the canonical theory is defined. The
desired Hilbert space is obtained by gluing together all
such tensor products (labelled by finite, ordered subsets of time
values) to form an infinite tensor product
$\otimes^\Omega_{t\in\mathR}\H_t$.

	Thus, in our version of the Gell-Mann and Hartle generalised
quantum theory, the basic ingredients are an orthoalgebra $\UP$ of
propositions about possible `histories' (or `universes'), and a space
$\D$ of decoherence functionals. A decoherence-functional is a
complex-valued function of pairs $\a,\b\in\UP$ whose value $d(\a,\b)$
is a measure of the extent to which the history-propositions $\a$ and
$\b$ are `mutually incompatible'. The pair $(\UP,\D)$ is to be
regarded as the history analogue of the pair $({\cal L},{\cal S})$ in
standard quantum theory where $\cal L$ is the orthoalgebra of
single-time propositions and $\cal S$ is the space of states on $\cal
L$. In this context we recall that the `or' operation (denoted $\oplus$)
in an orthoalgebra is defined only on pairs of `disjoint' elements,
whereas in a lattice the `or' operation $\vee$ in defined on any pair.
An orthoalgebra \cite{FGR92} seems to be the minimal structure that can
usefully be associated with the space $\cal L$ of single-time propositions,
although there have been many studies of the implications of assuming
that $\cal L$ is equipped with the additional operations of a lattice.

    It should be noted that the orthoalgebra of homogeneous and
inhomogeneous histories in standard quantum theory is actually a
proper subset of the family of all projection operators on
$\otimes^\Omega_{t\in\mathR}\H_t$, although it seems likely that the
remaining `exotic' histories can be generated from this preferred
subset by the application of the full lattice operations. In any
event, it is clearly of considerable interest to contemplate examples
of the Gell-Mann and Hartle scheme in which the space $\UP$ of all
history propositions is modelled by the projection lattice $\PV$ of
some Hilbert space $\V$ but where $\V$ has no {\em prima facie\/}
connection with tensor products of temporally-labelled spaces. In particular,
it is important to understand the structure of the associated space of
decoherence functionals.

    This requires a careful specification of what is meant by a
`decoherence functional', and how this differs from the more familiar
notion of a state. To this end,  recall that a state $\s\in\cal S$ on
an orthoalgebra $\cal L$ is defined to be a real-valued function on $\cal
L$ with the following properties:
    \be
    \item {\em Positivity\/}: $\s(P)\ge0$ for all $P\in\cal L$.

    \item {\em Additivity\/}: if $P$ and $R$ are disjoint then
          $\s(P\oplus R)=\s(P)+\s(R)$. This requirement
          is usually extended to include countable collections of
          propositions.

    \item {\em Normalisation\/}: $\s(1)=1$
    \ee
where $1$ is the unit proposition that is always true.

    The analogous properties for a decoherence
functional $d:\UP\times\UP\rightarrow\mathC$ are postulated to be:
    \be
    \item {\em Hermiticity\/}: $d(\a,\b)=d(\b,\a)^*$ for all
          $\a,\b$.
    \item {\em Positivity\/}: $d(\a,\a)\ge0$ for all $\a$.
    \item {\em Additivity\/}: if $\a$ and $\b$ are disjoint then, for
          all $\g$, $d(\a\oplus\b,\g)=d(\a,\g)+d(\b,\g)$. If
          appropriate, this can be extended to countable sums.
    \item {\em Normalisation\/}: $d(1,1)=1$.
    \ee

    The possibility of modelling history-propositions by elements of a
projection lattice $\PV$ raises the fundamental question of what the
space $\D$ is for an orthoalgebra of this type. The analogous question
in normal quantum theory is answered by Gleason's famous theorem
\cite{Gle57}: the states $\s$ on a projection lattice
$\P(\H)$ (with $\dim\H>2$) are in one-to-one correspondence with
density matrices $\rho$ on $\H$ with
    \beq
        \s_\rho(P)=\tr(P\rho){\rm\ for\ all\ } P\in\P(\H).  \label{Gleason}
    \eeq
 The main aim of the present paper is to prove the analogue of
Gleason's theorem for the history theory. Specifically, we shall show
that the decoherence functionals $d\in\D$ on a projection lattice
$\PV$ (with $\dim\V>2$) are in one-to-one correspondence with
operators $X$ on the tensor product $\V\otimes\V$ with
    \beq
        d_X(\a,\b)=\tr(\a\otimes\b X){\rm\ for\ all\ }\a,\b\in\PV
                                                        \label{main}
    \eeq
and where $X$ satisfies:
    \be
    \item $X^\dagger=MXM$ where the operator $M$ is defined on
          $\V\otimes\V$ by $M(u\otimes v):=v\otimes u$;

    \item for all $\a\in\PV$, $\tr(\a\otimes\a X_1)\geq0$ where
          $X=X_1+iX_2$ with $X_1$ and $X_2$ hermitian;

    \item $\tr(X_1)=1$.
    \ee

    The plan of the paper is as follows. In section \ref{Sec:standard}
we show how the decoherence functional \eq{Def:d} can be written in the
form $\eq{main}$. As well as motivating \eq{main}, this demonstration
shows why it is natural to represent a homogeneous history $\h{\a}$ by
a tensor product
$\a_{t_1}\otimes\a_{t_2}\otimes\cdots\otimes\a_{t_n}$. This provides
an alternative to the approach used in \cite{Ish94a} and \cite{IL94a}
which was based on ideas of temporal logic. The result \eq{main}
itself is proved in section \ref{Sec:Gleason}, and the paper ends with
a short discussion of how this approach might be used to
construct generalised quantum theories.

\section{The tensor-product form of a standard decoherence functional}
\label{Sec:standard}
We shall now show how the standard decoherence functional \eq{Def:d}
 \beq
    d(\a,\b)=\tr\big(\a_{t_n}(t_n)\a_{t_{n-1}}(t_{n-1})\ldots\a_{t_1}(t_1)
      \rho_{t_0}\b_{t_1'}(t_1')\b_{t_2'}(t_2')\ldots\b_{t_m'}(t_m')\big)
                                        \label{Def:dfull}
 \eeq
can be written in the `Gleason' form \eq{main}.

    Let $A_1,A_2,\ldots,A_n$ be a set of trace-class
operators on a Hilbert space $\H$ and consider the trace of their
product
 \beq
    \tr_\H(A_1A_2\ldots A_n)=\sum_{i_1}\bra{e^1_{i_1}}
                A_1A_2\ldots A_n\ket{e^1_{i_1}} \label{2dum1}
\eeq
where $\{e^1_{i_1}\bigm|i_1=1,2,\ldots,\dim(\H)\}$ is an orthonormal
basis for $\H$. Let $\{e^j_{i_j}\bigm|i_j=1,2\ldots,\dim(\H)\}$,
$j=2,3\ldots n$, be a collection of $n-1$ such basis sets, and use
them to write \eq{2dum1} as
\beq
    \tr_\H(A_1A_2\ldots A_n)=\sum_{i_1,i_2,\ldots,i_n}
    \bra{e^1_{i_1}}A_1\ket{e^2_{i_2}}
    \bra{e^2_{i_2}}A_2\ket{e^3_{i_3}}   \ldots
    \bra{e^n_{i_n}}A_n\ket{e^1_{i_1}}.    \label{2dum2}
\eeq
The definition of the inner product on the tensor product of
Hilbert spaces is such that
$\bra{\psi_1\otimes\psi_2}A\otimes B\ket{\phi_1\otimes\phi_2}
=\bra{\psi_1}A\ket{\phi_1}\bra{\psi_2}B\ket{\phi_2}$, and hence
\eq{2dum2} can be rewritten in the suggestive form
 \beqa
\lefteqn{\tr_\H(A_1A_2\ldots A_n)=\hspace{-1cm}\label{2dum3} }	\\
& &\hspace{-1cm}\sum_{i_1,i_2,\ldots,i_n}
\bra{e^1_{i_1}\otimes e^2_{i_2}\otimes\cdots\otimes e^{n-1}_{i_{n-1}}
                            \otimes e^n_{i_n}}
         A_1\otimes A_2\otimes\cdots\otimes A_n
    \ket{e^2_{i_2}\otimes e^3_{i_3}\otimes\cdots\otimes e^n_{i_n}
                            \otimes e^1_{i_1}}.	\nonumber
\eeqa

    The next step is to construct the linear operator
$S:\otimes^n\H\rightarrow \otimes^n\H$ defined on product vectors by
 \beq
    S(v_1\otimes v_2\otimes\cdots\otimes v_{n-1}\otimes v_n):=
      v_2\otimes v_3\otimes\cdots\otimes v_n\otimes v_1
 \eeq
 and then extended by linearity and continuity in the usual way to
give a unitary operator on $\otimes^n\H$ (it can easily be checked that
$S^\dagger(v_1\otimes v_2\otimes\cdots\otimes v_{n-1}\otimes v_n)=
(v_n\otimes v_1\otimes v_2\otimes\cdots\otimes v_{n-1})$.
The expression \eq{2dum3} can be rewritten using $S$ as
 \beq
  \tr_{\H}(A_1A_2\ldots A_n)=
    \tr_{\otimes^n\H}(A_1\otimes A_2\otimes\cdots\otimes A_n S)
                                                    \label{prodtotp}
 \eeq
 which shows the natural relationship between traces of operator
products and tensor products of operators.

    The result \eq{prodtotp} is the key to rewriting \eq{Def:dfull} in
the desired form \eq{main}. However, to do so requires a couple of
tricks. The first is that
$A_1\otimes A_2\otimes\cdots\otimes A_k\otimes\cdots\otimes A_n$ can
be written as
\beq
    A_1\otimes A_2\otimes\cdots\otimes A_k\otimes\cdots\otimes A_n=
    (A_1\otimes A_2\otimes\cdots\otimes 1_k\otimes\cdots\otimes A_n)
    (1_1\otimes 1_2\otimes\cdots\otimes A_k\otimes\cdots\otimes 1_n)
\eeq
where $1_k$ denotes the unit operator on the $k$'th Hilbert space.
Hence
\beqa
    \lefteqn{\tr_{\H}(A_1A_2\ldots A_n)}                \\ \nonumber
&=&\tr_{\otimes^n\H}\big\{
    (A_1\otimes A_2\otimes\cdots\otimes 1_k\otimes\cdots\otimes A_n)
        (1_1\otimes 1_2\otimes\cdots\otimes A_k\otimes
                        \cdots\otimes 1_n)S\big\}   \\ \nonumber
&=& \tr_{\otimes^n\H}\big\{
    (A_1\otimes A_2\otimes\cdots\otimes 1_k\otimes\cdots\otimes A_n)
        Y\big\}                                     \\ \nonumber
&=& \tr_{\otimes^{n-1}\H}\big\{
    (A_1\otimes A_2\otimes\cdots\otimes A_{k-1}\otimes A_{k+1}
            \otimes\cdots\otimes A_n) Y'\big\}
\eeqa
 where  $Y:=(1_1\otimes 1_2\otimes\cdots\otimes
A_k\otimes\cdots\otimes 1_n)S$, and $Y'$ is obtained from $Y$ by
tracing over a complete set of states for the $k$'th Hilbert space.
Thus the form \eq{prodtotp} is preserved under the action of removing
any element $A_k$ by partial tracing.

    The second observation is the following. Let $U_1,U_2,\ldots, U_n$
be a collection of unitary operators on $\H$, and define
$A_k^{U_k}:=U_kA_kU_k^\dagger$. Then
\beq
  A_1^{U_1}\otimes A_2^{U_2}\otimes\cdots\otimes A_n^{U_n}=
    (U_1\otimes U_2\otimes\cdots\otimes U_n)
      (A_1\otimes A_2\otimes\cdots\otimes A_n)
        (U_1^\dagger\otimes U_2^\dagger\otimes\cdots\otimes U_n^\dagger)
\eeq
and hence
\beqa
\lefteqn{\tr_{\H}(A_1^{U_1}A_2^{U_2}\ldots A_n^{U_n})}  \\ \nonumber
    &=&\tr_{\otimes^n\H}\big(
        A_1^{U_1}\otimes A_2^{U_2}\otimes\cdots\otimes A_n^{U_n}S\big)
     = \tr_{\otimes^n\H}\big(
        A_1\otimes A_2\otimes\cdots\otimes A_n S^U\big)
\eeqa
 where $S^U:=(U_1^\dagger\otimes U_2^\dagger\otimes\cdots\otimes
U_n^\dagger)S(U_1\otimes U_2\otimes\cdots\otimes U_n)$. Hence the
form \eq{prodtotp} is also preserved under the action of
an arbitrary unitary transformation on each operator $A_k$.

    Thus, using the above results to remove the initial density matrix
$\rho_{t_0}$ and the unitary time-evolution operators around each
Schr\"odinger-picture projection operator, it follows
that there exists some bounded operator $Z$ on $\otimes^{n+m}\H$ such that,
for all $\a,\b$, the decoherence functional \eq{Def:dfull} can be written as
 \beq
    d(\a,\b)=\tr_{\otimes^{n+m}\H}(
        \a_{t_n}\otimes\a_{t_{n-1}}\otimes\cdots\otimes\a_{t_1}\otimes
        \b_{t_1'}\otimes\b_{t_2'}\otimes\cdots\otimes\b_{t_m'}Z),
                                        \label{d_almost}
\eeq
which is almost in the desired form.

    The final step is to construct the linear `reversal' map
$R_{(n)}:\otimes^n\H\rightarrow\otimes ^n\H$ by defining it first on
product vectors as
\beq
    R_{(n)}(v_1\otimes v_2\otimes\cdots\otimes v_n):=
        v_n\otimes v_{n-1}\otimes\cdots\otimes v_1      \label{Def:opR}
\eeq
 and then extending to all vectors in the usual way. We note that $R_{(n)}$
is hermitian and satisfies $R^2_{(n)}=1$. When applied to a homogeneous
operator it gives $R_{(n)}(A_1\otimes A_2\otimes\cdots\otimes
A_n)R_{(n)}=A_n\otimes A_{n-1}\otimes\cdots\otimes A_1$. In particular,
 \beqa
\lefteqn{\hspace{-1cm}\a_{t_n}\otimes\a_{t_{n-1}}\otimes\cdots\otimes\a_{t_1}
       \otimes \b_{t_1'}\otimes\b_{t_2'}\otimes\cdots\otimes\b_{t_m'}
=R_{(n)}\otimes 1_{t_1'}\otimes 1_{t_2'}\otimes\cdots\otimes 1_{t_m'}\times}
                                                         \nonumber\\
    &&\hspace{-2cm}
(\a_{t_1}\otimes\a_{t_2}\otimes\cdots\otimes\a_{t_n}\otimes
        \b_{t_1'}\otimes\b_{t_2'}\otimes\cdots\b_{t_m'})
        R_{(n)}\otimes 1_{t_1'}\otimes 1_{t_2'}\otimes\cdots\otimes 1_{t_m'}
\eeqa
and hence
\beq
    d(\a,\b)=\tr_{\otimes^{n+m}\H}(
        \a_{t_1}\otimes\a_{t_2}\otimes\cdots\otimes\a_{t_n}\otimes
        \b_{t_1'}\otimes\b_{t_2'}\otimes\cdots\otimes\b_{t_m'}X),
                                        \label{d_Gleason}
\eeq
where
\beq
    X:=(R_{(n)}\otimes 1_{t_1'}\otimes 1_{t_2'}\otimes\cdots\otimes 1_{t_m'})
    Z(R_{(n)}\otimes 1_{t_1'}\otimes 1_{t_2'}\otimes\cdots\otimes 1_{t_m'}).
\eeq
 This concludes the demonstration of how a standard decoherence
functional \eq{Def:dfull} can be written in the Gleason form
\eq{d_Gleason}.

    Finally we note how the construction above can be extended to the
full tensor-product space $\V:=\otimes^\Omega_{t\in\mathR}\H_t$ where
$\Omega$ is a map from $\mathR$ to $\H$ with
$\braket{\Omega(t)}{\Omega(t)}=1$ for all $t\in\mathR$. The basic
homogeneous vectors in $\V$ are defined as maps $v:
\mathR\rightarrow\H$ such that $v(t)=\Omega(t)$ almost everywhere, \ie
for all but a finite number of values of $t$. The inner product on a
pair of such vectors is $\braket{v}{w}:=
\prod_{t\in\mathR}\braket{v(t)}{w(t)}$, which is well-defined since
$\braket{v(t)}{w(t)}=1$ for all but a finite number of $t$ values. The
infinite tensor product $\otimes^\Omega_{t\in\mathR}\H_t$ is defined
by taking the set of formal finite combinations of such vectors, then
identifying any pair whose inner product vanishes, and finally
completing the ensuing pre Hilbert space.

    A natural way of constructing operators on
$\V:=\otimes^\Omega_{t\in\mathR}\H_t$ is to take functions
$A:\mathR\rightarrow {\cal B}(\H)$ that are equal to the unit operator
almost everywhere. Any such function acts on a vector
$v:\mathR\rightarrow \H$ by $(Av)(t):= A(t)v(t)$ and preserves the
property of being equal to $\Omega$ almost everywhere. As such, it
extends to a well-defined operator on the completed Hilbert space
$\V$.

    In the case of the operators $S$ and $R_{(n)}$ defined above we proceed
as follows. First define $S$ on the vector $\Omega$ by
$S\Omega:=\Omega$. Now let $v:\mathR\rightarrow\H$ be a vector with
temporal support $\{t_1,t_2,\ldots,t_n\}$, \ie $v(t)=\Omega(t)$ for
all but this finite set of $t$-values, and define $Sv$ by
$(Sv)(t_i):=t_{i+1\,{\rm mod}\, n}$. Clearly $Sv$ differs from
$\Omega$ on the same set of $t$ values as does $v$, and hence it can
be extended to an operator on the full space
$\otimes^\Omega_{t\in\mathR}\H_t$. Likewise we define the reversal map
$R$ by $R\Omega:=\Omega$ and $(Rv)(t_i):=v(t_{n-i})$. Once again this
gives a well-defined operator on $\V$. By this means we are able to
extend the result $d(\a,\b)=\tr(\a\otimes\b X)$ to the set of all
history propositions in standard quantum theory .

\section{The Main Result}
\label{Sec:Gleason}
 The discussion that follows is restricted to the situation where
the Hilbert space $\V$ has a finite dimension (as did Gleason in the
original proof of his theorem on states) and we shall comment only
briefly on how to extend the theorem to the
infinite-dimensional case. We also require the decoherence functionals
$d:\PV\times \PV\rightarrow\mathC$ to be {\em continuous\/} functions
with respect to the natural topology on the space $\PV$ of projection
operators on $\V$. Under these conditions we can prove the following
analogue of Gleason's theorem for decoherence functionals, defined as
continuous maps $d:\PV\times \PV\rightarrow\mathC$ that satisfy the
four conditions listed in section \ref{Sec:intro}:
hermiticity, positivity, additivity, and normalisation.

\medskip
\noindent
{\bf Theorem}

\medskip
\noindent
If $\dim\V>2$, decoherence functionals $d$ are in one-to-one
correspondence with operators $X$ on $\V\otimes\V$ according to the
rule
\beq
        d(\a,\b)=\tr_{\V\otimes\V}(\a\otimes\b X)       \label{GLhist}
\eeq
with the restrictions that:
\beqa
    &a)&\ \tr_{\V\otimes\V}(\a\otimes\b X)=
        \tr_{\V\otimes\V}(\b\otimes\a X^\dagger)
            {\rm\ for\ all\ }\a,\b\in\PV,               \label{herm}\\
    &b)&\ \tr_{\V\otimes\V}(\a\otimes\a X)\ge0
            {\rm\ for\ all\ }\a\in\PV,                  \label{pos} \\
    &c)&\ \tr_{\V\otimes\V}(X)=1.                       \label{norm}
\eeqa

\bigskip
\noindent
{\bf Proof\/}

\noindent
One way round, the theorem is trivial. Namely, if a function $d$ is
{\em defined\/} by the right hand side of \eq{GLhist} it clearly
obeys the crucial additivity condition. The
extra requirements \eq{herm}, \eq{pos} and \eq{norm} then guarantee
hermiticity, positivity and normalisation respectively.

\smallskip
\noindent
 Conversely, let $d:\PV\times\PV\rightarrow\mathC$ be a decoherence
functional. The proof that it must have the form \eq{GLhist} exploits
Gleason's famous result on states on the lattice $\PV$.

    The first step is to define, for each $\a\in\PV$, the function
$d_\a:\PV\rightarrow\mathC$ where $d_\a(\b):=d(\a,\b)$. Let $\Re
d_\a$ and $\Im d_a$ denote the real and imaginary parts of $d_\a$, so
that
\beq
    d_\a(\b)=\Re d_\a(\b)+i\Im d_\a(\b)
\eeq
 with $\Re d_\a(\b)\in\mathR$ and $\Im d_\a(\b)\in\mathR$. Then the
bi-additivity condition on $d\in\D$ means that $\Re d_\a$ and $\Im
d_\a$ are additive functions on $\PV$, \ie $\Re d_\a(\b_1\oplus\b_2) =
\Re d_\a(\b_1)+\Re d_\a(\b_2)$ for any disjoint pair $\b_1,\b_2$ of
projectors; and similarly for $\Im d_\a$.

    Now $d(\a,\b)$ is a bi-continuous function of its arguments
$\a,\b\in\PV\times\PV$, and hence $\b\mapsto d_\a(\b)$ is a continuous
function on $\PV$, as are its real and imaginary parts. However the
set of all projectors in the finite-dimensional space $\V$ is a finite
disjoint union of compact Grassman manifolds, and is hence compact. It
follows that the functions $\b\mapsto\Re d_\a(\b)$ and $\b\mapsto\Im
d_\a(\b)$ are bounded below and above. On the other hand, for any
$r\in\mathR$, the quantity
 \beq
    \kappa_r(\b):=r\dim(\beta)=r\tr(\beta)
\eeq
 is a real additive function of $\b$, and hence so are $\Re
d_\a+\kappa_r$ and $\Im d_\a+\kappa_s$ for any $r,s\in\mathR$. Since
$\Re d_\a$ is bounded below, there exists $r_\a\in\mathR$ so that
$(\Re d_\a+\kappa_{r_\a})(\b)\ge0$ for all $\b$ (for example, choose
$r_\a:=|\min_{\b\in\PV}\Re d_\a(\b)|$), and similarly for each $\a$ there
is a real number $s_\a$ so that $(\Im d_\a+\kappa_{s_\a})(\b)\ge0$ for
all $\b\in\PV$. Furthermore, since $\Re d_\a+\kappa_{r_\a}$ and $\Im
d_\a+\kappa_{s_\a}$ are bounded above, we can choose positive real scale
factors $\mu_\a$ and $\nu_\a$ such that, for each $\a$ and for all $\b$
\beqa
        0&\leq&\mu_\a(\Re d_\a+\kappa_{r_a})(\b)\leq 1       \\
        0&\leq&\nu_\a(\Im d_\a+\kappa_{s_a})(\b)\leq 1.
\eeqa

    These inequalities plus the additivity property show that, for
each $\a\in\PV$, the quantities $\b\mapsto\mu_\a(\Re
d_\a+\kappa_{r_a})(\b)$ and $\b\mapsto\nu_\a(\Im
d_\a+\kappa_{s_a})(\b)$ are states on the lattice $\PV$. Then
Gleason's thereom shows that, for each $\a\in\PV$, there exists a
pair of density matrices $\rho^1_\a$ and $\rho^2_\a$ on $\V$ such that,
for all $\b\in\PV$,
 \beqa
        \mu_\a(\Re d_\a+\kappa_{r_\a})(\b)&=&\tr_\V(\rho^1_\a\b)\\
        \nu_\a(\Im d_\a+\kappa_{s_\a})(\b)&=&\tr_\V(\rho^2_\a\b),
\eeqa
and so
\beqa
    \Re d_\a(\b)&=&\tr_\V\big(({1\over\mu_\a}\rho^1_\a-r_\a)\b\big)=
                                \tr_\V(Y^1_\a\b)\\
    \Im d_\a(\b)&=&\tr_\V\big(({1\over\nu_\a}\rho^2_\a-s_\a)\b\big)=
                                \tr_\V(Y^2_\a\b)
\eeqa
 where $Y^1_\a:={1\over\mu_\a}\rho^1_\a-r_\a$ and
$Y^2_\a:={1\over\nu_\a}\rho^2_\a-s_\a$. Thus we have shown the
existence of a family of operators $Y_\a:=Y^1_\a+iY^2_\a$, $\a\in\PV$,
on $\V$ such that \beq
            d(\a,\b)=\tr_\V(Y_\a\b).                 \label{halfway}
\eeq

    The next step is to note that the additivity condition
$d(\a_1\oplus\a_2,\b)=d(\a_1,\b)+d(\a_2,\b)$ implies that
\beq
\tr_\V(Y_{\a_1\oplus\a_2}\b)=\tr_\V(Y_{\a_1}\b)+\tr_\V(Y_{\a_2}\b)
\eeq
 which, since it is true for all $\b\in\PV$ (and hence for all operators
on $\V$), implies that the operator-valued map $\a\mapsto Y_\a$ is itself
additive in the sense that
\beq
        Y_{\a_1\oplus\a_2}=Y_{\a_1}+Y_{\a_2}        \label{Yadd}
\eeq
 for all disjoint pairs of projectors $\a_1,\a_2$ on the Hilbert space
$\V$.

    Clearly what we need now is an analogue of Gleason's
theorem for operator-valued functions. Let
$\{B_i \bigm| i=1,2,\ldots,\dim(\V)^2\}$ be a vector-space basis for the
operators on $\V$, so that any operator $A$ can be expanded as
$A=\sum_{i=1}^{\dim(\V)^2} a_i B_i$ where $a_i\in\mathC$. In
particular, such an expansion can be made for the operators $Y_\a$
with $Y_\a=\sum_i y_i(\a)B_i$. Then the relation
\eq{Yadd} shows that the complex expansion-coefficients $y_i(\a)$,
$i=1,2,\ldots\dim(\V)^2$, must satisfy the additivity condition
\beq
        y_i(\a_1\oplus\a_2)=y_i(\a_1)+y_i(\a_2).
\eeq
 Now we invoke the same type of argument that lead to \eq{halfway} and
conclude the existence of operators $\Lambda_i$ on $\V$ such that
 \beq
    y_i(\a)=\tr_\V(\a\Lambda_i)
\eeq
and hence
\beq
        Y_\a=\sum_i\tr_\V(\a\Lambda_i)\,B_i.
\eeq
In particular,
\beq
   d(\a,\b)=\tr_\V\left\{\sum_i\big(\tr_\V(\a\Lambda_i)\big)B_i\b\right\}
                =\sum_i\tr_\V(\a\Lambda_i)\,\tr_\V(B_i\b). \label{3dum1}
\eeq

    The final step is to define an operator $X$ on $\V\otimes\V$ by
\beq
            X:=\sum_i\Lambda_i\otimes B_i
\eeq
and compute
\beq
    \tr_{\V\otimes\V}(\a\otimes\b X)=
        \sum_i\tr_{\V\otimes\V}(\a\otimes\b\, \Lambda_i\otimes B_i)=
        \sum_i\tr_\V(\a\Lambda_i)\tr_\V(\b B_i),
\eeq
which shows that \eq{3dum1} can be written in the desired form
\beq
    d(\a,\b)=\tr_{\V\otimes\V}(\a\otimes\b X)
\eeq
as claimed.

    This completes the proof of the theorem since the conditions
\eqs{herm}{norm} follow at once from the hermiticity, positivity, and
normalisation conditions on decoherence functionals.

    The conditions \eq{herm} and \eq{pos} can be expressed in a way
that is a little more transparent. We start by defining the reversal
map $M:\V\otimes\V\rightarrow\V\otimes\V$ by $M(u\otimes v):=v\otimes
u$, so that on operators $A\otimes B$ we have $B\otimes A=M(A\otimes
B)M$. Then condition \eq{herm} can be rewritten as
 \beq
    \tr_{\V\otimes\V}(\a\otimes\b X)=
        \tr_{\V\otimes\V}\big(M(\a\otimes\b)MX^\dagger\big)=
            \tr_{\V\otimes\V}(\a\otimes\b MX^\dagger M)
\eeq
which, since it is true for all projectors of the form $\a\otimes\b$,
implies that \eq{herm} is equivalent to the condition
\beq
            X^\dagger=MXM.                          \label{condXdag}
\eeq
If we now write $X=X_1+iX_2$ where $X_1$ and $X_2$ are hermitian,
then \eq{condXdag} is equivalent to the pair of conditions
\beqa
            X_1&=&  MX_1M                         \\    \label{condX1}
            X_2&=& -MX_2M.                              \label{condX2}
\eeqa

    These equations have an interesting implication for the positivity
requirement
\eq{pos} that $\tr_{\V\otimes\V}(\a\otimes\a X)\ge0$. The
posited reality of $\tr_{\V\otimes\V}(\a\otimes\a X)$ implies that
$\tr_{\V\otimes\V}(\a\otimes\a X_2)=0$ for all $\a$, which appears to
be quite a strong restriction. However, the relation $X_2=-MX_2M$
gives
 \beqa
    \tr_{\V\otimes\V}(\a\otimes\a X_2)&=&-\tr_{\V\otimes\V}(\a\otimes\a MX_2M)
						\\
        &=&-\tr_{\V\otimes\V}\big(M(\a\otimes\a)MX_2\big)=
			-\tr_{\V\otimes\V}(\a\otimes\a X_2)\nonumber
 \eeqa
 and so $\tr_{\V\otimes\V}(\a\otimes\a X_2)=0$ is implied already
by the hermiticity requirement \eq{condXdag}. Hence the only
independent conditions on $X$ are
 \beqa
        &&X^\dagger=MXM,                                     \\
        &&\tr_{\V\otimes\V}(\a\otimes\a X_1)\ge0
                                {\rm\ for\ all\ }\a\in \PV,
\label{concldum1}\\
        &&\tr_{\V\otimes\V}(X_1)=1.			\label{concldum2}
\eeqa
Note that \eq{concldum1} is considerably weaker than requiring $X_1$ to
be a positive operator on the Hilbert space $\V\otimes\V$. Indeed, in
\cite{IL94a} examples were given in normal quantum theory of inhomogeneous
histories $\a$ and decoherence functionals $d$ for which $d(\a,\a)>1$. Because
of
the normalisation condition \eq{concldum2} this can be
achieved only if the associated $X_1$ operator has some negative eigenvalues.

    This completes our discussion of the history analogue of Gleason's
theorem in the case where the Hilbert space $\V$ has a finite
dimension. It seems very likely that the theorem can be extended to an
infinite dimensional space by placing appropriate restrictions on the
decoherence functional $d$, probably some type of boundedness
condition on the complex numbers $d(\a,\b)$. We hope to return in a
later paper to this issue and to the related problem of classifying
decoherence functionals on non type-I von Neumann algebras.

\section{Conclusion}
\label{Sec:concl}
 Our proof of the history analogue \eq{main} of Gleason's theorem
opens up the possibility of constructing a wide range of examples of
the generalised quantum theory proposed by Gell-Mann and Hartle. The
classification of decoherence functionals is particularly important
because, as well as giving the putative probabilities of histories, a
decoherence functional also contains whatever dynamical, or
quasi-dynamical structure is present (of course, in a quantum theory
of gravity there may be none at all), as well as any initial
conditions.  Thus, for example, if the general representation
$d_X(\a,\b)=\tr(\a\otimes\b X)$ is applied to standard quantum theory,
the crucial `decoherence operator' $X$ depends on both the initial
density matrix $\rho_{t_0}$ and the Hamiltonian operator $H$ that
appears in the unitary evolution operator
$U(t,t')=e^{-i(t-t')H/\hbar}$. This situation contrasts strongly with
that of standard quantum theory where the state of the system
determines only the probabilities of single-time propositions:
dynamical evolution appears externally as a one-parameter family of
automorphisms of the lattice or orthoalgebra of such propositions.

    The work in \cite{Ish94a} and \cite{IL94a} showed that the space
$\UP$ of history propositions in standard quantum theory can indeed be
associated with the projection lattice of a certain tensor-product
Hilbert space $\V:=\otimes^\Omega_{t\in\mathR}\H_t$, and our
discussion in section \ref{Sec:standard} of the present paper throws
more light on why tensor products arise naturally in this context.
However, our expectation is that important instances of the general
scheme will exist in which the set $\UP$ of `propositions about the
universe' is modelled by a projection lattice $\PV$, but where the
vector space $\V$ has no connection with tensor products of
temporally-labelled Hilbert spaces.

    This raises the intriguing question of how such a vector space
$\V$ might arise. One can ask a similar question for the  Hilbert
space $\H$ of standard quantum theory whose lattice $\P(\H)$
represents the set of all propositions about the system at a single
time. One well-known answer is that $\H$ is the carrier of an
irreducible representation of the Lie group of the canonical
commutation relations; for example, for motion on the line $\mathR$
the Stone von Neumann theorem for the representations of the
Weyl-Heisenberg group $W$ shows that $\H$ can always be taken as
$L^2(\mathR)$.

    Suppose now we focus on $n$-time homogeneous histories and ask for
the origin of the Hilbert space $L^2_{t_1}(\mathR)\otimes
L^2_{t_2}(\mathR)\otimes\cdots\otimes L^2_{t_n}(\mathR)$. One answer,
given in \cite{Ish94a,IL94a}, is to invoke ideas of quantum temporal
logic. Another possibility is the one discussed in section
\ref{Sec:standard} of the present paper where we demonstrated the
natural connection between tensor products and traces of products of
operators. However, a different way of looking at this issue is to say
that $\otimes^nL^2(\mathR)$ arises as the carrier of an irreducible
representation of the {\em product\/} $\prod_{i=1}^{i=n}W_i$ of $n$
copies of the canonical Weyl group $W$. This idea of constructing the
history theory by a `temporal-gauging' of the canonical group was
mentioned briefly in \cite{Ish94a} but it becomes far more important
now we have some insight into the structure of decoherence functionals
on a projection lattice.

    Thus the general idea is that a vector space $\V$ whose projectors
in $\PV$ represent history/universe propositions can be obtained by
looking for representations of a `history-group' $\cal G$ which serves
as an analogue of the temporally-gauged canonical group of standard
quantum theory. This idea places the problem of constructing
generalised history theories within an overall mathematical scheme
that involves three well-defined steps: i) find a history group $\cal
G$; ii) study its irreducible representations on Hilbert spaces $\V$;
and iii) use the ideas in the present paper to explore the decoherence
functionals for this system.

    Of course, it is not trivial to identify the history group $\cal
G$ for a particular system, but this problem is not dissimilar to the
more familiar problem of finding the appropriate canonical group with
which to construct a canonical quantisation of a given classical
system. The study of decoherence functionals is also more complicated
than the analogous study of states since, as remarked above, whatever
quasi-dynamical structure there may be is coded in the choice of
decoherence functional. However, in general terms this does seem to be
a very promising approach to constructing generalised quantum theories
of the Gell-Mann and Hartle type, and we plan to give some concrete
examples in forthcoming papers.

\bigskip
\bigskip
\noindent
{\large\bf Acknowledgements}

\noindent
We gratefully acknowledge support from the SERC and Trinity
College, Cambridge (CJI), and the Leverhulme and Newton Trusts (NL).
SS has been supported
by a DAAD fellowship HSPII financed by the German Federal Ministry of Research
and Technology. We are all grateful to the staff of the Isaac Newton Institute
for providing such a congenial working environment.

\end{document}